# Superconducting dome and field-enhanced superconductivity of PLD synthesized $Nd_{1-x}Eu_xNiO_2$ thin films

Wenlong Yang[1,2†], Qiang Zhao[1,3†], Xingke Fu[1†], Gaofei Ren[1,3], Zhongjing Wu[1,3], Zhen Chen[1,3], Jianping Sun[1,3], Boseng Wang[1,3], Jiacai Nie[2,4*], Pengtao Yang[1,3*], Jinguang Cheng[1,3*]

[1]*Beijing National Laboratory for Condensed Matter Physics and Institute of Physics, Chinese Academy of Sciences, Beijing, 100190 China*

[2]*School of Physics and Astronomy, Beijing Normal University, Beijing 100875, People's Republic of China*

[3]*School of Physical Sciences, University of Chinese Academy of Sciences, Beijing 100049, China*

[4]*Key Laboratory of Multiscale Spin Physics, Ministry of Education, Beijing Normal University, Beijing 100875, People's Republic of China*

*E-mail: jcnie@bnu.edu.cn; ptyang@iphy.ac.cn; jgcheng@iphy.ac.cn*

We report on the synthesis of infinite-layer $Nd_{1-x}Eu_xNiO_2$ ($0 \leq x \leq 0.7$) thin films using pulsed laser deposition (PLD) followed by topotactic reduction with $CaH_2$. Resistivity measurements on these films reveal a superconducting dome within the doping range $0.2 \leq x \leq 0.5$, which is wider than that achieved by molecular beam epitaxy and comparable to that obtained by chemical synthesis. The $x = 0.3$ PLD film exhibits the optimal superconducting transition temperature $T_c \approx 31$ K, much higher than those grown by other vacuum epitaxial techniques. This result indicates that PLD is an ideal approach for fabricating high-quality, high-$T_c$ $Nd_{1-x}Eu_xNiO_2$ superconducting films. Magneto-transport measurements reveal robust field-enhanced and re-entrant superconductivity in both underdoped and overdoped regimes. At low temperatures just above the onset $T_c$, the Hall resistance exhibits nonlinear behavior, which may originate from magnetic impurity scattering. These results highlight the crucial role of magnetic rare-earth $Eu^{2+}$ ions in producing the exotic physical properties of the infinite-layer nickelates.

## Introduction

The infinite-layer nickelate is obtained by removing the apical oxygen from the perovskite structure via low-temperature topotactic reduction[1-4]. Following the initial discovery of superconductivity in $Nd_{0.8}Sr_{0.2}NiO_2$[1], the introduction of magnetic rare-earth ions, particularly $Eu^{2+}$, has emerged as a new route to tune both the lattice and electronic degrees of freedom [5-11]. In (Sm-Eu-Ca-Sr)$NiO_2$ thin films, the superconducting transition temperature $T_c$ has been elevated to nearly 40 K. It was later found that the Eu-doped infinite-layer nickelate thin films with lower $T_c$ exhibit field-induced enhanced or reentrant superconductivity upon application of an external magnetic field—phenomena previously observed in organic, Chevrel phase, and heavy-fermion superconductors [9,12-16]. These field-induced behaviors have been largely attributed to the Jaccarino-Peter (J-P) effect, i.e., the paramagnetic polarization of localized $Eu^{2+}$ moments by external field generates an internal exchange field that partially compensates the applied external magnetic field. Depending on the doping level and $T_c$, the competition between this exchange field and the intrinsic upper critical field leads to either field-enhanced or re-entrant superconductivity[7,9,10,17,18]. Notably, it has been suggested that the J-P effect alone cannot fully account for the observations in Ref. [17], implying a possible mechanism of magnetically mediated superconductivity. Thus, the Eu-doped infinite-layer nickelates offers an ideal platform for investigating the interplay between rare-earth magnetism and unconventional superconductivity.

Despite these advances, preparing high-quality, pure Eu-doped infinite-layer nickelate films without alkaline-earth co-doping remains challenging, especially via pulsed laser deposition (PLD). For example, PLD-grown $Sm_{0.8}Eu_{0.2}NiO_2$ film exhibits a $\boldsymbol{T}_c$ of only 7.8 K. Three distinct approaches have been reported for synthesizing pure Eu-doped $Nd_{1-x}Eu_xNiO_2$ thin films across a wide doping range: molecular beam epitaxy (MBE) with in-situ Al reduction yields a relatively narrow superconducting dome ($0.15 < x < 0.4$ ) with $\boldsymbol{T}_c < 30$ K [19]; off-axis RF magnetron sputtering followed by Al reduction gives a lower $T_c$ even for the optimally doped $Nd_{0.7}Eu_{0.3}NiO_2$ films [11] ; and the chemical solution method achieves a broader superconducting dome ($0.2 < x < 0.6$) with $\boldsymbol{T}_c$ exceeding 30 K [10]. Intriguingly, although field-enhanced and re-entrant superconductivity are commonly observed, the reported phase diagrams differ considerably in the optimal doping level, superconducting range, and the specific regions where re-entrance occurs. This discrepancy strongly indicates that extrinsic factors tied to the fabrication process, such as microscopic defects, doping homogeneity, and interface states, can profoundly modulate the coupling between rare-earth magnetic moments and superconducting order parameter.

In this letter, we report the successful PLD growth of $Nd_{1-x}Eu_xNiO_2$ thin films on $(LaAlO_3)_{0.3}(Sr_2TaAlO_6)_{0.7}$ (LSAT) substrate followed by $CaH_2$ topotactic reduction, and present the superconducting phase diagram of infinite-layer $Nd_{1-x}Eu_xNiO_2$ thin films. The PLD-prepared films exhibit a wider superconducting dome than those grown by MBE or sputtering, comparable to that obtained via chemical synthesis, and optimally doped PLD films also show a higher $T_c$ than other vacuum epitaxial methods, establishing PLD as an effective approach for expanding the doping window. Magneto-transport measurements reveal robust field-enhanced and re-entrant superconductivity in both underdoped and overdoped regions of the dome. Furthermore, Hall effect measurements uncover a pronounced nonlinear Hall response in the low-temperature non-superconducting regime, accompanied by negligible magnetoresistance hysteresis, pointing to magnetic impurity scattering. These results not only validate the PLD route for fabricating pure Eu-doped nickelate superconducting thin films but also provides fresh experimental insight into the intricate mechanisms that couple rare-earth magnetism and superconductivity.

## Results and Discussion

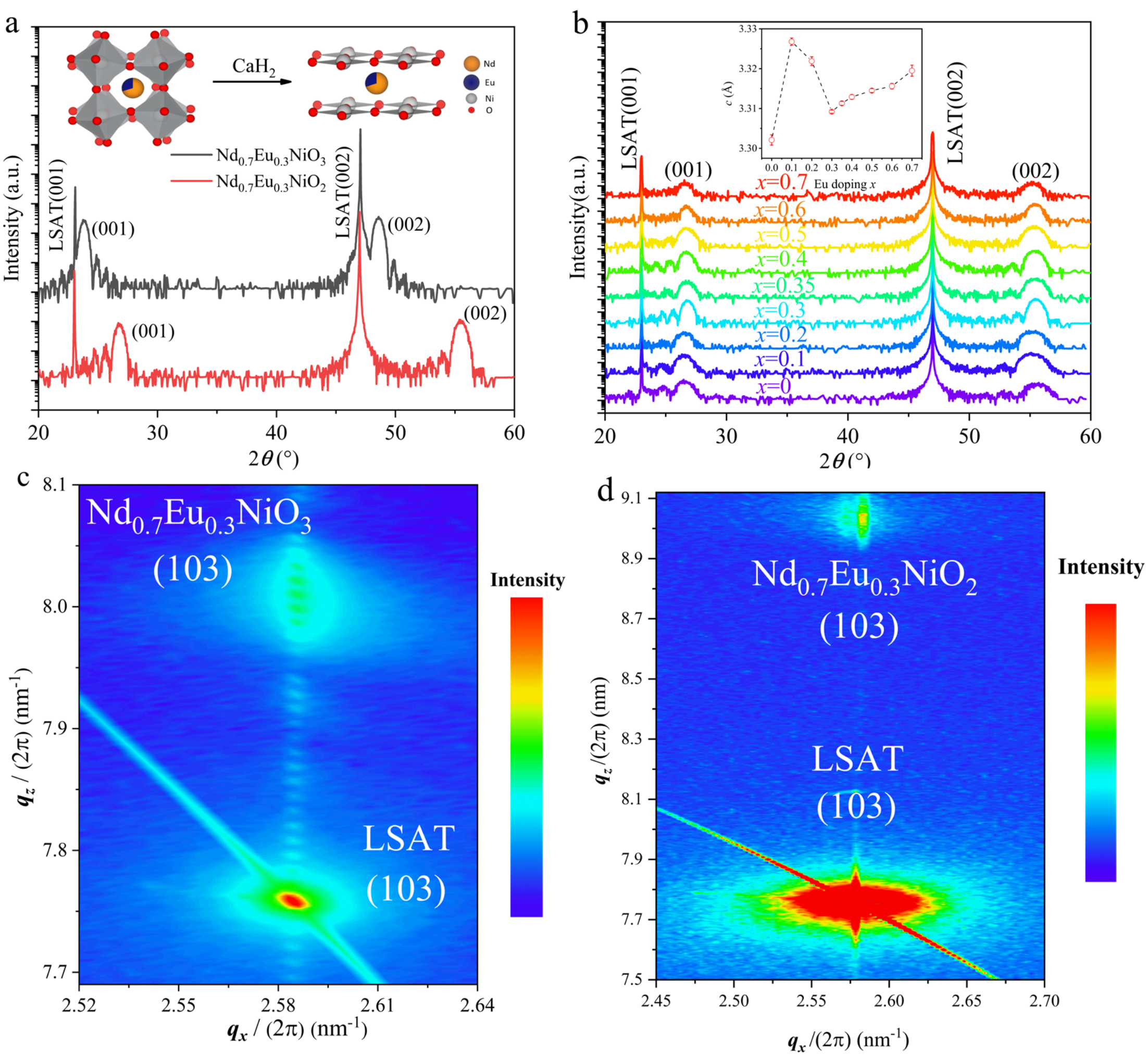

**Fig. 1 Structural characterization of $Nd_{1-x}Eu_xNiO_2$ thin films.** (a) XRD $\theta$–$2\theta$ scans of the $Nd_{0.7}Eu_{0.3}NiO_3$ precursor film and the $Nd_{0.7}Eu_{0.3}NiO_2$ film obtained after $CaH_2$ reduction. (b) XRD patterns of $Nd_{1-x}Eu_xNiO_2$ films for Eu doping levels $x$ = 0 to 0.6. The inset shows the room-temperature $c$-axis lattice constant $d$ as a function of Eu content $x$. (c, d) Reciprocal space maps of the $Nd_{0.7}Eu_{0.3}NiO_3$ and $Nd_{0.7}Eu_{0.3}NiO_2$ films.

Figure 1(a) shows the XRD $\theta$–$2\theta$ scans of the $Nd_{0.7}Eu_{0.3}NiO_3$ perovskite precursor film and the same film after $CaH_2$ reduction. The (001) and (002) reflections of the precursor at 23.82° and 48.59° shift to 26.82° and 55.49° upon reduction, confirming the successful formation of infinite-layer $Nd_{0.7}Eu_{0.3}NiO_2$. The XRD patterns for the full doping series $Nd_{1-x}Eu_xNiO_2$ ($x$ = 0-0.7) films are displayed in Fig. 1(b). The $c$-axis lattice constant, extracted from the (002) peak positions and plotted in the inset, varies non-monotonically with x, exhibiting an anomaly in the underdoped regime. We ascribe this behavior to an enhanced density of Ruddlesden-Popper (R-P) stacking faults, as confirmed by Scanning Transmission Electron Microscopy (STEM) measurements (see Supplementary Materials Fig. S1). Reciprocal space maps around the (103) reflection, Figs. 1(c) and (d), reveal that the films are epitaxially coherent with the LSAT substrate and remain fully strained both before and after reduction.

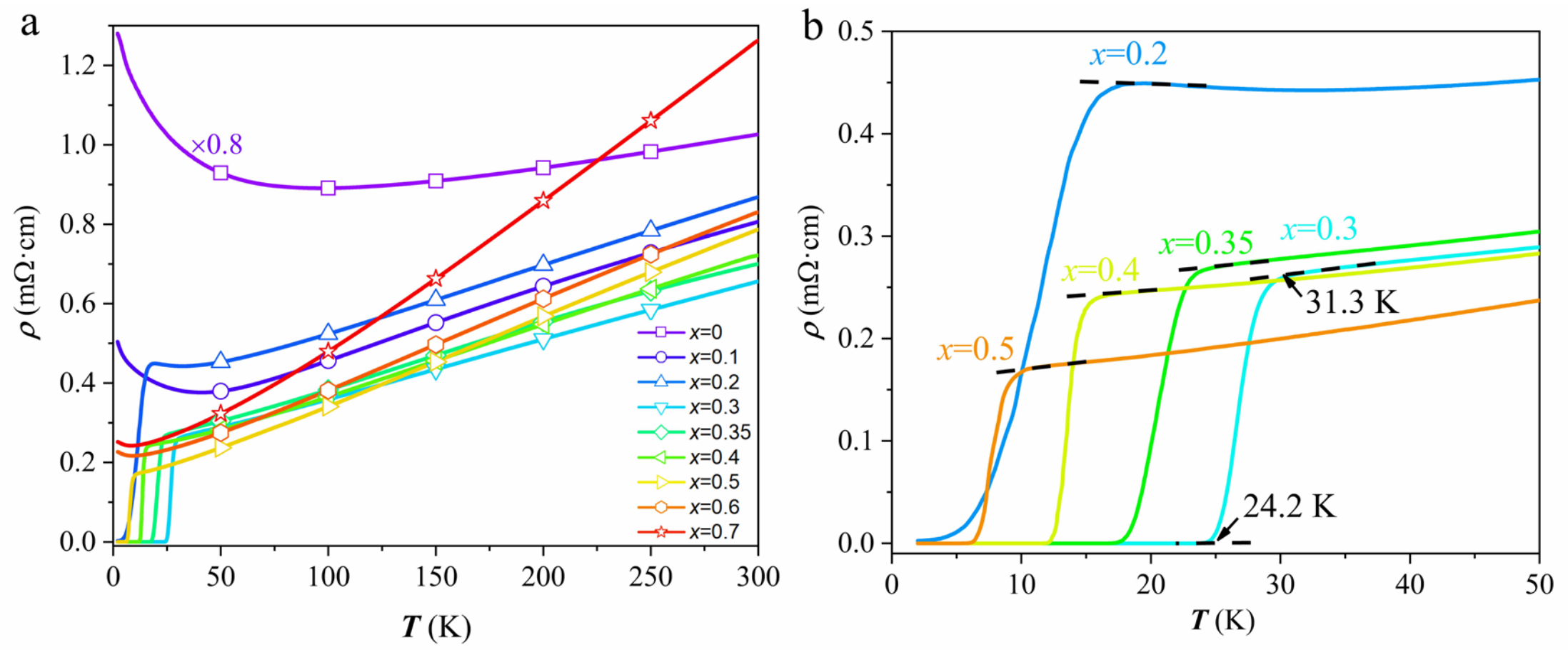

**Fig. 2 Electrical transport properties of $Nd_{1-x}Eu_xNiO_2$ films.** (a) Temperature dependence of resistivity $\rho(T)$ for $x$ = 0-0.7. (b) Enlarged view of the low-temperature (2-50K) $\rho(T)$ curves for superconducting films with $x$ = 0.2-0.5.

Figure 2(a) displays the resistivity $\rho(T)$ of $Nd_{1-x}Eu_xNiO_2$ films for $x$ = 0-0.7. In the underdoped regime ($x$ = 0, 0.1), a low-temperature resistivity upturn signifies weak insulating behavior, which is progressively suppressed with increasing Eu doping content. Superconductivity first emerges at $x$ = 0.2, when an upturn persists just above the onset $T_c$. For $x$ = 0.3-0.5, zero resistance is achieved and the normal-state $\rho(T)$ exhibits purely metallic character without any upturn. Figure 2(b) provides an enlarged view of the superconducting transitions in $\rho(T)$ for $x$ = 0.2-0.5. The onset (zero-resistance) temperatures $T_c^{onset}$ ($T_c^{zero}$) are 18.2 K (none), 31.3 K (24.2 K), 24.6 K (17.3 K), 16.7 K (12 K), and 10.6 K (6.1 K) for $x$ = 0.2, 0.3, 0.35, 0.4, and 0.5, respectively. For $x$ = 0.6 and 0.7, the transport behavior resembles that of the underdoped region, with a resistivity minimum shifted to a lower temperature of ~9 K. Such low-temperature weakly insulating behavior has been attributed to localization effects, Kondo physics[20], or disorder[21] according to previous studies on Sr-overdoped counterparts. To further substantiate the emergence of superconductivity, the critical current density ($J_c$) was measured for $x$ = 0.3 film at various temperatures (see Supplementary Materials Fig. S2). Notably, $\boldsymbol{J}_c$ reaches ~31.6 kA/cm$^2$ at 20 K, surpassing those reported for Sr-doped[1,22] infinite-layer and the bilayer R-P $La_2PrNi_2O_7$ [23] films.

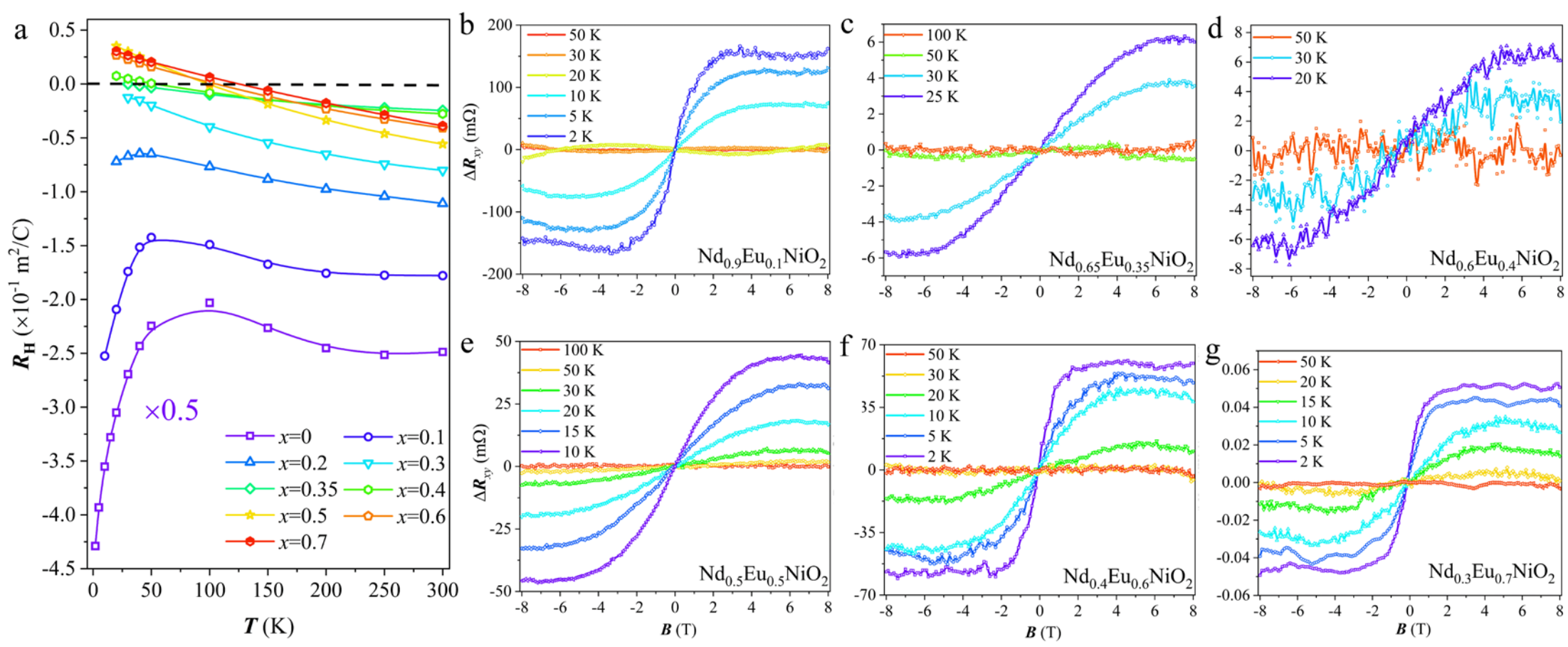

**Fig. 3 Hall effect in $Nd_{1-x}Eu_xNiO_2$.** (a) Temperature dependence of the Hall coefficient $R_H$ for different Eu doping levels x. The black dashed line marks the sign-change boundary. (b)-(g) Non-linear Hall resistance $\Delta R_{xy} = R_{xy}^{exp} - R_{xy}^{L}$ as a function of temperature for $x = 0.1$, 0.35, 0.4, 0.5, 0.6 and 0.7.

The Hall coefficients $R_H(T)$ is extracted from a linear fit to the Hall resistance $R_{xy}(B)$ over the full magnetic field range. Figure 3(a) shows $R_H(T)$ as a function of doping x and temperature. The magnitude of $R_H$ decreases progressively with increasing doping level, and a sign reversal sets in for $x > 0.3$, precisely coincidence with the overdoped region. In contrast to Eu-doped Sm-based infinite-layer nickelates[17], the sign-reversal temperature of $R_H$ increases linearly with $x$, pointing to a distinct Fermi surface evolution in the Nd-based system at high Eu doping.

As depicted in Figs. 3(b–g), nonlinear Hall transport appears in both the underdoped and overdoped regimes. The nonlinear Hall resistance is defined as $\Delta R_{xy} \equiv R_{xy}^{exp} - R_{xy}^{L}$, where $R_{xy}^{L}$ is obtained from the linear extrapolation of the high-field (6–8 T) region. For all doping levels, the saturation magnitude of $\Delta R_{xy}$ increases monotonically with decreasing temperature. No appreciable hysteresis is observed in either $R_{xx}$ or $R_{xy}^{exp}$ in the low-temperature normal state (see Supplementary Materials Fig. S3). Such nonlinear hall response is consistent with the findings of Yang et al.[17], who observed magnetic hysteresis inside the superconducting state, hinting at an unconventional coexistence of superconductivity and ferromagnetism. Varbaro et al.[11] attributed the nonlinear Hall resistivity in $Nd_{0.7}Eu_{0.3}NiO_2$ to the magnetic moments of $Eu^{2+}$ and $Nd^{3+}$ ions. Early studies on paramagnetic systems established that magnetic impurity scattering can also generate an anomalous Hall voltage[24-28]. The microscopic origin of the anomalous Hall effect in Eu-doped infinite-layer nickelates therefore remains an open question and requires further investigations.

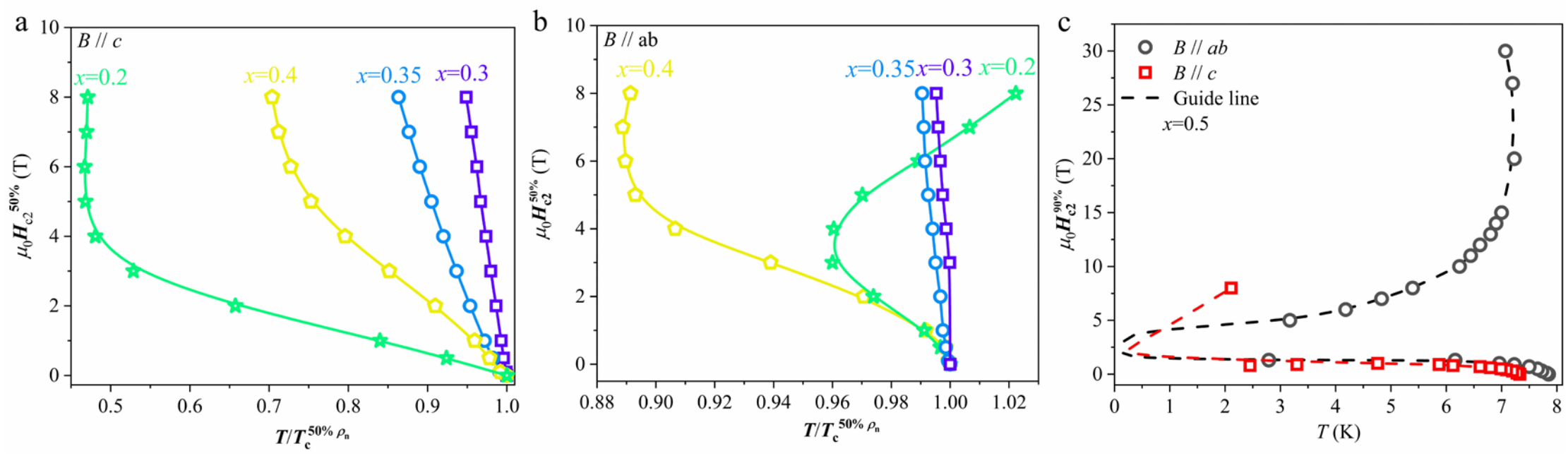

**Fig. 4 Upper critical field of $Nd_{1-x}Eu_xNiO_2$.** (a) Out-of-plane and (b) in-plane upper critical fields $\mu_0 H_{c2}$ as functions of reduced temperature $T/T_c^{50\%\rho_n}$ for $x = 0.2$-$0.4$, where $T_c^{50\%\rho_n}$ is defined by the 50% resistivity criterion. (c) Temperature dependence of $\mu_0 H_{c2}$ for $x = 0.5$; the dashed lines are guides to the eye.

Figure 4(a) and (b) present the out-of-plane ($B$ // $c$) and in-plane ($B$ // $ab$) upper critical fields $\mu_0 H_{c2}$ as functions of the reduced temperature $T/T_c^{50\%\rho_n}$ for $x = 0.2$, 0.3, 0.35 and 0.4 films. For the $x = 0.2$ and 0.4 films, a subtle kink appears near 4 T in both field orientations, signaling the onset of field-enhanced superconductivity. In contrast, the $x = 0.3$ and 0.35 films show no discernible kink features up to 8 T. Notably, for $Nd_{0.8}Eu_{0.2}NiO_2$ in an in-plane field, the kink at ~ 3T is pronounced and the superconducting transition temperature $T_c$ at 8 T even exceeds the zero-field value. This indicates that the J-P effect alone cannot fully explain the field-enhanced superconductivity, pointing to additional mechanisms for this phenomenon. Figure 4(c) shows $\mu_0 H_{c2}(T)$ for $x = 0.5$. In both orientations, superconductivity is strongly suppressed at low fields, while a pronounced re-entrant superconducting state emerges at high fields, reflecting a clear separation between these two regimes. Thus, we find two types of distinct field-induced phenomena in the series of $Nd_{1-x}Eu_xNiO_2$ films, i.e., field-enhanced superconductivity for x = 0.2, 04 and field-induced re-entrant superconductivity for x = 0.5.

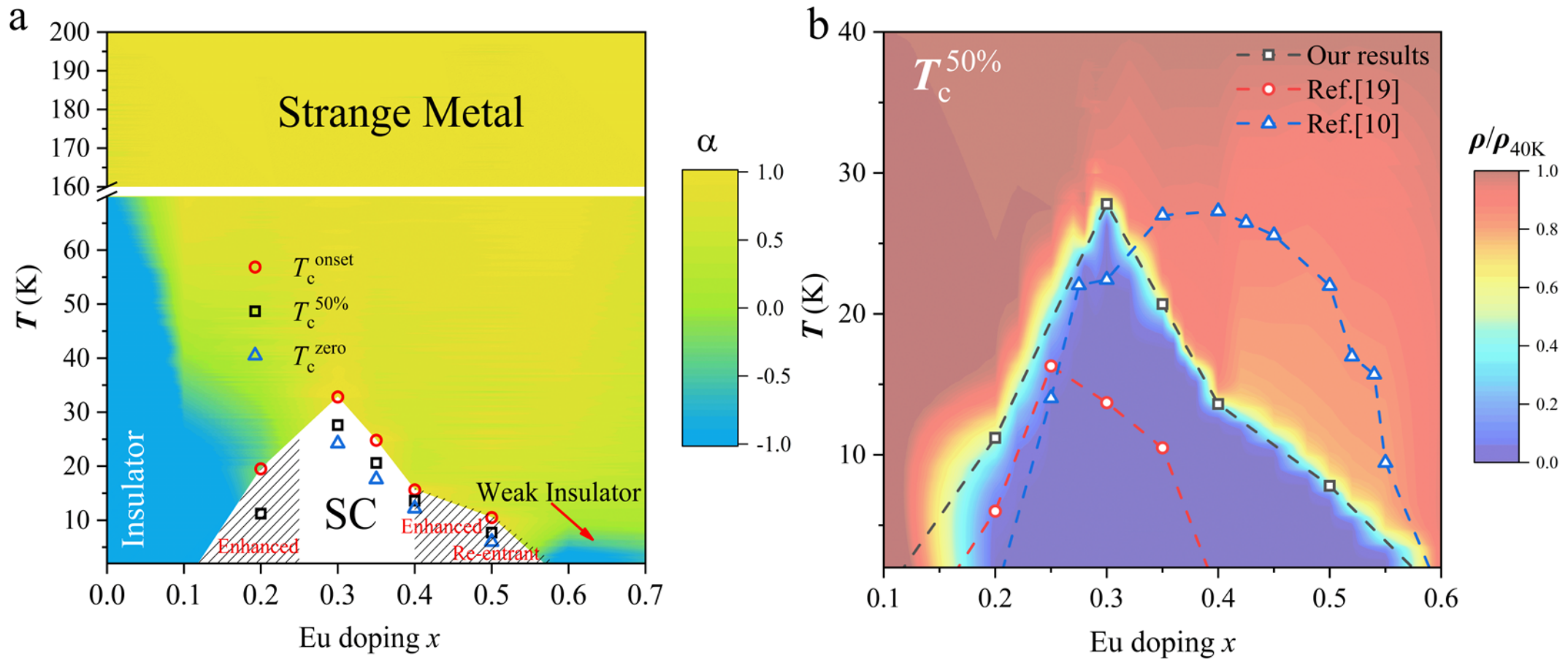

**Fig. 5 Phase diagram of $Nd_{1-x}Eu_xNiO_2$.** (a) Doping dependence of the superconducting transition temperature superimposed by a color contour plot of the resistivity exponent $\alpha = \partial\rho/\partial_T$, for which blue indicates insulating behavior ($\alpha < 0$) and yellow indicates strange metal behavior ($\alpha = 1$). Field reduced enhanced and re-entrant superconductivity appears in underdoped and overdoped regimes. (b) Comparison of the superconducting domes for the $Nd_{1-x}Eu_xNiO_2$ films grown by different methods; dashed lines are guides to the eye. The background color map displays the normalized resistivity $\rho(T)/\rho(40\ K)$ for the PLD films in this work.

Figure 5 presents the doping-dependent phase diagram of $Nd_{1-x}Eu_xNiO_2$ films. As shown in Fig. 5(a), field-enhanced superconductivity appears on both the underdoped and overdoped sides of the superconducting dome, whereas re-entrant superconductivity only occurs at the edge of the overdoped region (x = 0.5). In comparison to the Sr-doped system[21], the Eu-doped infinite-layer nickelate films exhibit a substantially extended strange-metal regime, which persists up to the overdoped boundary. This underscores the decisive role of the dopant species in producing the linear-in-$T$ resistivity. The overdoped region ($x > 0.4$) features an asymmetric tail in the superconducting phase boundary, suggesting the presence of a possible hidden quantum critical point.

Fig. 5(b) complies the $T_c^{50\%\rho_n}$ values obtained by different growth methods. As can be seen, the PLD films exhibit a much wider superconducting dome than those grown by MBE and comparable to those obtained via chemical synthesis. But the doping level for achieving optimal $T_c$ for the PLD films has been substantially shifted to lower x than those grown by chemical synthesis. In addition, the PLD films reach an optimal $T_c^{50\%\rho_n}$ of ~30 K at $x = 0.3$, substantially higher than that of MBE-grown films and comparable to the value achieved by the chemical solution route. These comparisons established PLD as an effective approach for fabricating high-quality, high-$T_c$ $Nd_{1-x}Eu_xNiO_2$ films with an extended doping window.

Regarding the disparity in the $\boldsymbol{T}_c$ among different thin films synthesis methods[7,10,11], we attribute the phenomenon to the following factors: Firstly, the discrepancy in topotactic reduction plays an important role. The in-situ Al reduction, while preventing surface contamination, is fundamentally limited to a solid-solid reaction. In contrast, $CaH_2$ reduction relies on the highly reactive hydrogen gas, which is more effective to remove the apical oxygens, leading to a more pristine infinite-layer structure. Secondly, the role of high oxygen pressure annealing cannot be overlooked. Both the fabrication of high-quality PLD targets and the chemical route require annealing under high oxygen pressure, which thermodynamically stabilizes the perovskite precursor.

Regarding the magnetotransport properties, previous studies reported the absence of field-enhanced superconductivity in the underdoped regime of (Sm,Eu,Ca,Sr)$NiO_2$ films[17]. Conversely, in $Nd_{1-x}Eu_xNiO_2$ thin films, field-enhanced superconductivity occurs in both the underdoped and overdoped regions, and field re-entrant superconductivity emerging at the overdoped edge[7,10]. We propose that the incorporation of alkaline-earth metals (Ca and/or Sr) as hole dopants in the (Sm,Eu,Ca,Sr)$NiO_2$ system shifts the entire superconducting dome toward lower Eu doping concentrations. Consequently, compared to the purely Eu-doped system, this composite system exhibits superconductivity at a markedly lower Eu concentration ($x$=0.06)[6] and reaches optimal doping at $x$=0.2[17]. Based on the J-P compensation effect, the total effective field experienced by the conduction electrons is given by $\boldsymbol{H}_T = \boldsymbol{H}_{ext} + \boldsymbol{H}_J$, where $\boldsymbol{H}_{ext}$ is the externally applied magnetic field and $\boldsymbol{H}_J$ is the exchange field generated via antiferromagnetic coupling. In the (Sm,Eu,Ca,Sr)$NiO_2$ system, the relatively low concentration of $Eu^{2+}$ (which provides the local moments) results in a weak exchange field $\boldsymbol{H}_J$. This weak $\boldsymbol{H}_J$ is insufficient to effectively compensate for the pair-breaking effect of the external field at lower fields, thereby failing to induce field-enhanced superconductivity. Nevertheless, it cannot be entirely ruled out that there are intrinsic differences between $Sm^{3+}$ and $Nd^{3+}$ regarding their roles in paramagnetically driven superconducting enhancement[11], which needs further experimental verification.

## Conclusion

In conclusion, we have successfully synthesized $Nd_{1-x}Eu_xNiO_2$ thin films by pulsed laser deposition. Optimal doped $Nd_{0.7}Eu_{0.3}NiO_2$ films exhibit an onset $\boldsymbol{T}_c$ exceeding 30 K, higher than those achieved by MBE and magnetron sputtering and comparable to the chemical route. Superconductivity emerges over a broad doping range of $0.2 \leq x \leq 0.5$, encompassing the windows of two prior methods and the defining the doping phase diagram for $Nd_{1-x}Eu_xNiO_2$. Hall measurements reveal a sign reversal of the Hall coefficient at $x > 0.3$ and nonlinear Hall resistance appears across a wide doping range at low temperatures, likely originating from scattering by the local magnetic moments. Significantly, field-enhanced and re-entrant superconductivity are observed on both the underdoped and overdoped sides of the superconducting dome. For the $x = 0.2$ sample, $T_c$ under an 8 T in-plane magnetic field surpasses the zero-field value, demonstrating that the Jaccarino-Peter effect alone cannot fully explain the field-enhanced superconductivity in this system. Our work establishes PLD as an effective route for synthesis of high-quality Eu-doped infinite-layer nickelates and underscores the profound role of magnetic rare-earth ions in their exotic superconducting properties.

## Experimental details

**Target preparation**. High-purity polycrystalline targets are essential for the growth of high-quality Eu-doped thin films. The perovskite nickelate is metastable and cannot be obtained by conventional solid-state reaction of NiO, $Nd_2O_3$, and $Eu_2O_3$; sintering under ambient pressure succeeds only when alkaline-earth Ca is incorporated. $Nd_{1-x}Eu_xNiO_3$ targets were then synthesized via a sol-gel route. $Nd(NO_3)_3 \cdot 6H_2O$, $Eu(NO_3)_3 \cdot 6H_2O$, and $Ni(NO_3)_2 \cdot 6H_2O$ (Aladdin, 99.99% purity) were dissolved in deionized water together with citric acid and nitric acid. The mixture was stirred in a water bath at 90 °C for ~4 hours to form a bright green nitrate gel, which was subsequently heated at 800 °C overnight to remove residual organics. The product was ground, pressed into pellets, and sintered at 1000 °C for 24 h under 80 bar $O_2$. Further details are reported in Ref. [29].

**Thin film growth**. The perovskite precursor films were deposited on LSAT (001) substrate by PLD using a 248 nm KrF excimer laser. The substrate temperature was held at 500–650 °C in an oxygen partial pressure of 20–30 Pa. The laser fluence was 2–3 J/cm$^2$ with a repetition rate of 4 Hz. After growth, the films were cooled to room temperature at 30 °C/min in the same oxygen atmosphere.

**Topotactic reduction.** The infinite-layer $Nd_{1-x}Eu_xNiO_2$ phase was obtained by a low-temperature soft-chemical reduction method. The precursor films were wrapped in clean Al foil and sealed in a quartz tube together with 0.15 g $CaH_2$ powder (Aladdin, 98% purity). The quartz tube was then evacuated to a base pressure of 0.1 Pa. Reduction was carried out at 290 °C for 2 h with both heating and cooling ramps of 10 °C/min.

**Transport measurements.** For zero-field and low-field electrical transport measurements (up to 9 T), a dry-variable-temperature superconducting magnet system from Physike Technology Co., Ltd. was employed. Resistivity measurements were performed using a standard four-probe method with Al wire-bonding. The Hall effect was measured using a standard six-point probe configuration to determine the Hall voltage. To investigate the electrical transport properties under high magnetic fields, high-field magnetoresistance measurements were conducted at a user station within the Synergetic Extreme Condition User Facility (SECUF) in Beijing. This station is equipped with a hybrid magnet system and capable of generating steady magnetic fields up to 35.6 T.

**Scanning transmission electron microscopy**. STEM specimens were prepared by a standard focused ion beam (FIB) lift-out procedure on a Thermo Fisher Helios 5CX. High-angle annular dark-field STEM (HAADF-STEM) imaging was performed on a spherical-aberration-corrected JEOL JEM-ARM200F NEOARM operated at 200 kV with a collection semi-angle of 90–370 mrad. Electron energy-loss spectra (EELS) data were recorded on the same instrument using a Gatan K3 camera, and energy-dispersive X-ray spectroscopy (EDS) mapping was carried out in STEM mode with a single EDS detector.

## References

[1] D. F. Li, K. Lee, B. Y. Wang, M. Osada, S. Crossley, H. R. Lee, Y. Cui, Y. Hikita, and H. Y. Hwang, Nature **572**, 624 (2019).
[2] S. W. Zeng et al., Phys. Rev. Lett. **133**, 7, 066503 (2024).
[3] P. P. Balakrishnan et al., Phys. Rev. Mater. **10**, 034801 (2026).
[4] W. Wei, K. Shin, H. Hong, Y. Shin, A. S. Thind, Y. Yang, R. F. Klie, F. J. Walker, and C. H. Ahn, Phys. Rev. Mater. **7**, 013802 (2023).
[5] S. L. E. Chow, Z. Luo, and A. Ariando, Nature **642**, 58 (2025).

[6] M. Yang et al., Nat. Commun. **17**, 2761 (2026).
[7] D. Vu et al., Nat. Commun. **17**, 3480 (2026).
[8] M. Yang et al., 2025), p. arXiv:2508.14666.
[9] K. Rubi et al., 2025), p. arXiv:2508.16290.
[10] H. Han et al., 2025), p. arXiv:2511.22026.
[11] L. Varbaro et al., Nat. Commun. (2026).
[12] S. Uji, H. Shinagawa, T. Terashima, T. Yakabe, Y. Terai, M. Tokumoto, A. Kobayashi, H. Tanaka, and H. Kobayashi, Nature **410**, 908 (2001).
[13] L. Balicas et al., Phys. Rev. Lett. **87**, 067002 (2001).
[14] C. Rossel, H. W. Meul, M. Decroux, O. Fischer, G. Remenyi, and A. Briggs, Journal of Applied Physics **57**, 3099 (1985).
[15] H. W. Meul, C. Rossel, M. Decroux, O . Fischer, G. Remenyi, and A. Briggs, Phys. Rev. Lett. **53**, 497 (1984).
[16] S. Khim et al., **373**, 1012 (2021).
[17] M. Yang et al., Nature **653**, 1052 (2026).
[18] L. Varbaro et al., 2026), p. arXiv:2601.19473.
[19] W. Z. Wei, D. Vu, Z. Zhang, F. J. Walker, and C. H. Ahn, Sci. Adv. **9**, 6, eadh3327 (2023).
[20] S. W. Zeng et al., Phys. Rev. Lett. **125**, 7, 147003 (2020).
[21] K. Lee et al., Nature **619**, 288 (2023).
[22] Q. Zhao et al., Phys. Rev. Lett. **133**, 6, 036003 (2024).
[23] Y. D. Liu et al., Nat. Mater. **24** (2025).
[24] C. M. Hurd and J. E. A. Alderson, Solid State Commun. **9**, 309 (1971).
[25] A. Fert and O. Jaoul, Phys. Rev. Lett. **28**, 303 (1972).
[26] A. Fert and A. Friederich, Phys. Rev. B **13**, 397 (1976).
[27] A. Fert, A. Friederich, and A. Hamzic, J. Magn. Magn. Mater. **24**, 231 (1981).
[28] A. Hamzić, S. Senoussi, I. A. Campbell, and A. Fert, J. Magn. Magn. Mater. **15-18**, 921 (1980).
[29] M. T. Escote, V. B. Barbeta, R. F. Jardim, and J. Campo, Journal of Physics: Condensed Matter **18**, 6117 (2006).